\documentstyle[12pt]{article}
\newcommand{\erwin}{Schr\"odinger}
\newcommand{\beq}{\begin{equation}}
\newcommand{\eeq}{\end{equation}}

\def\nuc#1#2{\relax\ifmmode{}^{#1}{\protect\text{#2}}\else${}^{#1}$#2\fi}
\begin{document}

\begin{center} {\Large\bf  A new perspective on quantum mechanics?}\\[5 mm]

{R.S. Mackintosh}\\[5 mm]
{\it School of Physical Sciences, The Open University, Milton Keynes, MK7 6AA, UK}\\[5mm]
\end{center}

{\bf Abstract:} {Bell suggested that a new perspective on quantum mechanics was needed. We propose a solution of the measurement problem based on a reconsideration of the nature of particles. The solution is presented with an idealized model involving non-locality or non-separability, identified in 1927 by Einstein and  implicit in the standard interpretation of single slit (or hole) diffraction.
Considering particles as being localizable entities leads to an `induced collapse model', a parameter-free alternative to spontaneous collapse models, that affords a new perspective on, \emph{inter alia}, nuclear decay.}

\begin{center}\today\end{center}
\section{Reporting outcomes of experiments}
There is no general agreement concerning the interpretation of quantum mechanics, QM.  John Bell, reflecting on this said~\cite{JSB} : "It may be that what we need is just some small change in the point of view, and everything will fall into a coherent whole, but it is extremely frustrating and also extremely interesting that we have not yet found this slight change in perspective." A new point of view is offered here.

There is no agreement as to how the formalism of QM is related to what, as Bohr pointed out, must be expressed in ordinary (classical) language  when the results of experiments are reported. Even the word `measurement' is problematic. 
John Bell, in his article `Against Measurement'~\cite{bellonmeasure},  was right to deplore the way the way the word `measurement' is applied in quantum physics. Here I note in particular that many natural measurement-like processes occur with no participation of human agency. Such a natural process is the creation of pleochroic haloes as described below. This is a process, exhibiting arguably the key characteristics of a measurement, that has been occurring since long before the appearance of humans. Accounts of measurement are often undermined by the fact that the word `measurement' in common usage suggests pre-existing properties.

We first consider an idealized experiment that might be performed in the absence of gravity. A nucleus of \nuc{238}{U} is levitated alone at the center of an evacuated (hence no cloud tracks) sphere the inner surface of which can detect swift charged particles by leaving a spot. After a wait, the nucleus will undergo  $\alpha$  decay. This can be described in terms of an S-wave ($l=0$) representing an entangled pair, a nucleus of $^4$He and a recoiling nucleus of \nuc{234}{Th}. The conventional description of alpha decay is in terms of a wave function for the relative $^4$He -- \nuc{234}{Th} distance in which the centre of mass coordinates do not appear. The entanglement is formally more apparent when the wave function is expanded as a sum of products of $\alpha$-particle and \nuc{234}{Th} coordinates; a basis in which the $\alpha$-particle is in a quite well-defined direction, is possible. For a useful detailed account of the entanglement of two-particle systems, in the context of the hydrogen atom,  see Refs.~\cite{Qva,PTom}.

A spot will appear on the detecting surface as the alpha particle hits it. It can then be predicted with certainty that a second spot will soon afterward appear on the exact opposite location on the spherical detecting surface. The second spot records the arrival of the thorium nucleus. This whole process places a natural event in a very artificial setup, but it is, in essence, something that occurs everywhere and all the time. It is not insignificant since, as Feynman has noted~\cite{boom}, such events make history strictly unpredictable.  But what occurs in this staged situation  involves the essence of what is called `measurement' in the context of quantum mechanics. Somehow, the transformation of an S-wave (more strictly, something \emph{described} by an S-wave) into a spot (strictly, a highly localized region where a detecting surface has changed its physical nature), a process sometimes called an `expansion into the macroscopic world', has taken place. There has not been a human measurer in sight. Yet, we commonly read, in defence of Qbism, for example, that a measurement can be understood in terms of the change in the understanding of a human observer. Thus  `measure'  has becomes confused with measuring by humans.

The S-wave is isotropic  and, moreover, the \nuc{238}{U} ground state nucleus has zero spin, thus having no spatial orientation. Thus alpha particles from many such nuclei at the centre of the sphere would leave a statistically isotropic set of spots. Each associated recoiling \nuc{234}{Th} would always hit the screen in the exact opposite position to the corresponding alpha spot.

Some might argue that the spots only appear much later when a person inspects the detecting surface of the sphere. Such an argument would also imply that the pleochroic haloes, which are found~\cite{pleo,ERuth,Hevesy} surrounding uranium- or thorium-bearing inclusions in  mica, only occur when the mica is examined in the laboratory. Such haloes, of course, are literally as old as the hills. 

Discussions of measurement in quantum mechanics should treat \emph{all} situations that involve the process that is commonly described as the collapse of the wave packet, or expansion into the macroscopic domain. Most such situations involve no human activity and appear to involve a departure from unitary evolution of a wave packet. Whatever the deeper ontology that might eventually emerge, the idea of a wave packet collapsing is a fair way of characterizing the transformation of something represented by an entangled S-wave, of $4\pi$ angular reach, into two spots at precisely  opposite sides of a spherical detecting surface.

Arguably, we need a new word to replace `measurement'. Perhaps `actualization' or `actualize' would indicate  something that the \emph{system} does rather than something a \emph{person} does, as is suggested by `measurement'. It is unclear to this author how spontaneous collapse models lead to actualization in the instant that the particle wave function encounters the detecting sphere, leaving, in the present case, two spots on exact opposite points on a sphere. A resolution is proposed in Section~\ref{spont}.

\section{Two slits or one?}
We now argue that Einstein's 1927 single hole experiment (Ref.~\cite{BV} and see below)  gets closer to the central mysteries of quantum mechanics than Feynman's two slits.  Feynman famously declared that all the mysteries of quantum mechanics are exposed in the two slit experiment. In introducing that experiment, he said: `I will take just this one experiment, which has been designed to contain all the mysteries of quantum mechanics, to put you up against the paradoxes and mysteries and peculiarities of nature one hundred percent.'~\cite{feynman1}. He went on to say: `Any other situation in nature, it turns out, can always be explained by saying ``remember the case of the experiment with the two holes? It's the same thing.'' '~\cite{feynman1} He pointed out that the electrons or photons behave as waves in that they give rise to a diffraction pattern, and as particles when they each leave a single spot on the detecting screen; in his words: `Electrons behave in this respect in exactly the same way as photons; they are both screwy, but in exactly the same way.'~\cite{feynman2} We agree with `both screwy in exactly the same way',  but question the unique role of the two slit experiment.

Many didactic accounts of quantum mechanics introduce the two slit example, and it is certainly  a fine illustration of a fundamental and characteristic quantum property, the superposition of amplitudes and the need to add \emph{amplitudes} before squaring. It also leads to a discussion of wave particle duality. According to the so-called Copenhagen interpretation, attributed to Bohr (but widely applied in a form due more to Heisenberg~\cite{who}), wave-like and particle-like properties are complementary The two slit experiment exemplifies this. If you ask where a single electron or photon arrives, you get a particle-like answer: at a single place. But the distribution of places of individual electrons or photons is wave-like. The two-slit diffraction pattern corresponds to interference between the amplitudes for going through one slit or the other. Wave-like and particle-like properties appear in a single experiment.

But is it really true that the two slit problem contains,  `... all the mysteries of quantum mechanics ...'?
Electrons, photons, or Bucky balls are still described by a wave packet even if one slit is closed. So, is one slit diffraction just as good an exemplar of complementarity? Imagine moving the two slits, step by step, closer together: the diffraction pattern will change continuously as the slits approach each other.  Eventually, they will become a single slit, and the particles will appear according to the laws of single slit diffraction. But, from the present point of view, nothing deep has changed. There are still particles (appearing) and still waves (determining their distribution). The single slit, or hole, experiment was actually done in 1910 by Taylor~\cite{taylor} with very low intensity light; the diffraction pattern appeared, spot by spot, with light of an intensity that we can now interpret as `one photon at a time'.

It was the single slit (or single hole) case that was the basis of Einstein's declaration in 1927, see Ref.~\cite{BV}, that quantum mechanics is \emph{incomplete};  unless, that is, some interpretation like that of de Broglie (later the de Broglie-Bohm (dBB)  theory) was applied. When de Broglie's interpretation dipped out of fashion, Einstein persisted in his belief in the incompleteness of quantum mechanics. Thus EPR~\cite{epr} devised their famous argument, involving \emph{two} particles, to demonstrate this incompleteness. Bell~\cite{bell} later showed that their model implied just what Einstein most abhorred: non-locality or non-separability. Ironically, for two particles the later dBB theory is also highly non-local, and itherefore does not solve Einstein's problem.  

Thus, it is single slit diffraction that most directly exhibits the deepest mysteries of QM \emph{pace} Feynman: non-locality and (at least apparent) wave function collapse. Regarding collapse, the wave packet for the single particle leaving the slit or hole (of whatever ontology) no longer exists after the spot appears on the screen. That applies to both `wave packet' in the logical category of elements of a formalism, or in the category of whatever physical entities are described by that formalism. The non-locality implicit in the conventional (collapse) model of single slit diffraction does not permit faster than light communication. 

The nonlocality of single slit diffraction is clear: according to Einstein~\cite{AE}, `...the real factual situation of the system $S_2$ is independent of what is done with system $S_1$, which is spatially separated from the former'. This is often referred to as Einstein locality. In the present context, if $S_1$ is a small region of a detecting screen, then the appearance of a spot at $S_1$ instantaneously forbids the particle to appear at any other separate region, $S_2$, wherever that may be on the screen. Einstein locality fails.

Although such nonlocality provides a clear  interpretation of single slit diffraction of particles, the most widely accepted evidence for non-locality involves more than one particle.  For a review of the non-locality revealed by the many EPR-type experiments involving two or more particles, see Ref.~\cite{brunner}. The two slit experiment remains valuable for teaching the superposition principle. In fact, as Jammer points out~\cite{jammer-p}, the double slit initially appeared to be a stumbling block for Born's interpretation.

Another quote from Feynman is relevant: `A philosopher once said ``It is necessary for the very existence of science that the same conditions always produce the same result''. Well, they do not.'~\cite{feynman3}. But there is an important qualification: if many people do the same diffraction experiment, with the same hole, the same screen, with particles having the same mass and energy, they would get (with \emph{many} particles) the same diffraction pattern. There is a case to say that this qualifies as realism, but a realism of quantum entities, propensities. So maybe the serious question now is `How identical are identically prepared states?' In any case, the existence of science contimues when the statistical results are replicated.

\section{A new approach to the measurement problem}
Calling a photon a particle may seem odd; a photon seems nothing like an electron, which we all agree is a particle. The position coordinates of an electron appear in Schr\"odinger's equation while those of a photon do not appear in Maxwell's equations. Moreover, an electron  has a definite location in its rest frame, even if this position is indeterminate in line with the uncertainty principle; a photon has no rest frame in which to have a definite position. Advocates of the dBB interpretation point to the evident particleness of electrons, declarimg this to be a strength of that interpretation. But the difficulty of dBB to treat photons in the way it treats electrons is a problem. For a discussion of photons treated as elementary particles see Ref.~\cite{BBBB}.

But there \emph{is} a perspective from which electrons and photons do share the essence of particle-ness. I refer to the perspective in which a particle is \emph{exactly} an entity that declares one place (within the resolution of the detector) when we ask where it is.  Photons and electrons share this property; it's part of their being `screwy in exactly the same way' as Feynman said. This places photons and  the other spin-one bosons, W and Z particles, in related ontological categories.  

From this perspective, the collapse of the wave function in a position measurement is a key aspect of the nature of a particle, suggesting that collapsability is a \emph{defining} property of particles rather than wavepackets. The theory that governs their propagation up to the point where the wave packet meets a detector is standard quantum mechanics. Following the first interaction of the particle which `collapses' its momentum state, the subsequent track in a cloud chamber would be determined by predominantly forward scattering, as explained by Mott~\cite{mott} long ago. In the case of alpha decay of a $^{238}$U nucleus embedded in a crystal lattice, the alpha particle will most probably have a well defined momentum from early on when the $^{234}$Th nucleus, that is entangled with it in an S-wave, interacts with the lattice.

Is an alpha particle, with four nucleons, a `particle' in the above sense? Yes. It certainly leaves a single spot on a detecting screen etc. Within any composite particle made of fermions, such as an alpha particle, the individual fermions are strongly entangled. The simplest possible structure will be a Slater determinant. The characteristic of an entangled state is that the individual particles do not each have a complete independent set of properties. We conclude from this that a particle which is a bound state of fermions acts as a unit, with a non-local cooperative response. Thus, the collapse of the wave packet of a propagating alpha particle is no more surprising than electrons or photons leaving spots when interacting with a screen, or the non-local collapse of an entangled pair in an EPR-type experiment.

Not everything qualifies as a particle in the present sense; no-one believes that cats can be diffracted. For diffraction,  a composite particle must have a  well-defined Hamiltonian with a specified number of sub-particles (electrons, nucleons...). It must also \emph{not} be entangled with the environment. A breathing, hair-shedding feline is disqualified on both counts, and the well publicised `cat states' of AMO physics cannot  be particles representative of \erwin's cats. \erwin's cat fable simply shows that unitary evolution is not the whole story.

An important consequence of our suggested characterization of particle-ness follows if, as has been proposed, any measurement is in essence a position measurement; this is certainly the case for a spin measurement in a Stern-Gerlach experiment. As Bell~\cite{SUS} remarked: `\ldots in physics the only observations we must consider are position observations.' If this is universally true, then the collapse at the core of the measurement problem can be viewed not just as a wave function collapse but, as  a consequence of the defining property, as proposed here, of a particle. Bell also implied the significance of position measuremendt in his discussion of beables in QFT: `What is essential is to be able to define the positions of things ...' , see Bell~\cite{bellbeables}.

The picture presented here does not, so far, explain how an electron, for example, presents as a particle at a particular place. A full account is required, not least, by cosmology. Perhaps the best clue regarding an electron reaching the detecting screen, is that the electron's inelastic interaction with the screen interrupts its state of motion, triggering a chain of events. These events would  rapidly reach a threshold such that they could not take place elsewhere in the screen.  The probability of the process appearing elsewhere is vanishingly small in the same way as the likelihood of all the gas molecules appearing on one side of a container. In this light the `measurement' event  is not a \emph{spontaneous localization}, but an \emph{induced localization}.   

The characterization of `particle' proposed here supports a realistic interpretation in which probabilities amount to propensities of the particles to appear in accordance with the Born postulate. This makes the concept of `identically prepared states' a key element of QM. A basic postulate is the supposition that an immediate second measurement of some observable leads to the same result.  It is therefore reasonable  to consider, in principle, truly identically prepared states.  If the general uncertainty principle, GUP, as derived by Robertson~\cite{Robertson} had appeared before Heisenberg's heuristic argument, in which the disturbance by a `measurer' is implicitly present, the subsequent history of the interpretation of QM might have been quite different. It might also have been different if Born had referred to frequencies rather than probabilities.

The  GUP involves objective uncertainties that are defined in terms of expectation values based on identically prepared states. These objective uncertainties can be calculated in a simple way, but they could also be measured in a mechanized way without intervention of a human measurer as part of the measurement. That is, unless the original designer of the equipment, which we suppose general enough to apply in a variety of cases, can be considered a measurer. But it is un-natural to consider the instrument designer to be disturbing the system in the Heisenberg sense. Of course, the key enabling concept behind expectation values is that of  \emph{identically prepared states}, something that is not always sufficiently stressed.

\section{Implications} The word `particle' implies `non-wave' but implies something that appears in a single place when `actualized', and this suggests the language we use to describe particle diffraction should be modified.  In the two-slit case, rather than speak of a \emph{particle} going through one slit or another, refer to an \emph{electron}, if that is what it is, reaching the slits. With a single slit, the wave packet of an electron propagates through the slit \ldots unless it doesn't because the wave packet has `collapsed', i.e. the electron has actualized, on the screen with the slit. In that case the electron's encounter with the screen with the slit is effectively a `measurement' revealing that its position, as it reached the screen with the slit, is never   \emph{within} the slit. A spot on the screen with the slit would already reveal the particle nature of an electron. An electron is an entity that presents as a particle when it leaves a spot. If, however, the electron does reach the detecting screen, this would not only contribute to a single slit diffraction pattern, but also constitute a measurement of the position of the particle incident on the screen with the slit: it was never actualized on the screen with the slit, outside the region of the slit.

In the two-slit case, leading after the passage of many electrons to a two-slit diffraction pattern, the electrons that produce that pattern will each have passed through both slits. To describe this there is no need for `particle-or-wave' agonizing: an electron is an entity whose propagation is described by a wave, and whose appearance is in a spot, a property suggesting `particle' in a classical context. It is better to speak of an \emph{electron} going through the two slits, in the case of a double slit, and avoid the confusing language of a \emph{particle} going through one slit, or the other slit, or both; its particle nature will be manifest when it hits a position sensitive detector. In short, in answer to the question `is it a particle or a wave?' is: it is an electron (or other specific particle) whose propagation is described by a wave and whose particle nature is revealed when it is detected.

The above picture implies that the electrical charge of a diffracted electron is not localized until the `actualization' when it leaves a spot. Hence, we may also view the wave aspect of bound particles differently since the wave nature of an electron is not confined to its propagation as a free particle. Consider a helium atom in a Rydberg state in which one electron is in a ground state (strictly, the two electrons are entangled so neither is wholly `ground state' or wholly `Rydberg state'). In this (approximate) picture, the highly excited Rydberg electron will `see' a spherically symmetric charge density related to $|\psi|^2$ for the ground state electron. Likewise, in the classic Nobel prize experiments of Hofstadter~\cite{hofs}, which measured the size of atomic nuclei, the very high energy electrons were deflected by a smooth charge distribution due to the nucleons. From the pattern of the deflection of the high energy electrons, the radial charge densities of nuclei were deduced. In fact, according to of Bohr's complementarity, the answer you get depends on the question you ask, so for example, ultra high-energy (short wavelength) incident electrons might reveal sub-nucleonic structure.

\section{Induced collapse model}\label{spont} Spontaneous collapse models have been put forward to solve the measurement problem. They involve unknown parameters and do not explain apparent collapses that are not considered spontaneous, such as the appearance of a spot that appears just when a particle hits the screen. The spontaneous aspect does appear when a particle is emitted by a radioactive nucleus. 

An alternative approach exploits the notion of a `particle' as a localizable entity, rather than as a local entity. Consider again the alpha particle decay of a $^{238}$U nucleus. According to the account given in elementary quantum text books, the isotropic stationary state wavefunction  of the alpha particle outside the nuclear Coulomb barrier will be \emph{very} small; so small that it takes of the order of $2 \times 10^7$ years for a single such nucleus to make an impression on the environment. Historically this might be a faint scintillation in a spinthariscope or a track in mica~\cite{pleo,ERuth,Hevesy}. A scintillation will be accompanied non-locally by a transition of the $^{238}$U nucleus to a nucleus  of $^{234}$Th.  There was no \emph{spontaneous} collapse of the entangled pair, but an \emph{induced} collapse of  the entangled pair as the mass-4 member of the pair interacts irreversibly within absorptive material and the spherical character of the alpha `particle' disappears. Now the mass four entity is unarguably a `particle.'  Alternatively, as  pointed out above, the actualization could already have come about when the entangled recoiling nucleus was stopped. Either way, the probability of a decay occuring is predominantly a function of the magnitude of the radial wavefunction for the members of the entangled pair and is equally likely in all directions consistently with its angular momentum quantum number of zero. 

\section{Final remarks} The licence for non-locality, granted by the EPR-type experiments that were inspired by Bell's  inequalities, and a key to the present interpretation, has far-reaching consequences~\cite{shimony}. Non-locality was already considered deeply disturbing by Einstein very early in the history of quantum mechanics, as empahsized by Howard~\cite{DH-AE}, who notes:
\begin{quotation}Here, then, we have Einstein on the eve of the 1927 Solvay meeting. For twenty-two years he had known that the full story of the quantum would involve some fundamental compromise with the classical notion of the mutual independence of interacting systems. In 1924, Bose showed him that this failure of mutual independence was not incidental but an essential feature of the quantum realm, the deep fact underlying the Planck formula.\end{quotation} 
Einstein was consistent in his desire to eliminate non-locality, or non-separability,  from physics. This led to the EPR paper with the eventual outcome: the establishment of non-locality as an experimental fact.  Together with the status of position measurements, this supports a simple interpretation of `measurement' in terms of a macroscopic event, an `actualization', involving a defining property of `particle': \emph{A particle is an entity which appears at a single position when its location is established in a macroscopic event.} The actualization is a nonlocal process that applies to photons, electrons, nuclei, and (to date) quite large molecules, but not to cats or other objects that have no quantum states independent of the environment. Rather than being a \emph{localized} entity (desired by dBB), a particle is a \emph{localizable} entity. It makes possible an \emph{induced collapse model} which seems more natural than spontaneous collapse models.

Einstein in 1927 was entirely justified in being unhappy with a particle, propagating as a wave, instantaneously disappearing at all but a single spot when encountering a detecting screen, a non-local process. Bell's analysis, applied to the many increasingly loophole-free EPR-type experiments, provides strong evidence for non-locality~\cite{brunner} of a specific kind. The non-locality apparent, but seldom acknowledged, when identical wave packets actualize as diffraction patterns, is therefore entirely plausible. It is not the least remarkable aspect of quantum mechanics.  The licence for non-locality must also be a key element in applying the proposed nature of particles to a 3N dimensional entangled wave packet.

\end{document}